# Stable Facts, Relative Facts

Andrea Di Biagio[1] · Carlo Rovelli[2,3,4]

## Abstract

Facts happen at every interaction, but they are not absolute: they are relative to the systems involved in the interaction. Stable facts are those whose relativity can effectively be ignored. In this work, we describe how stable facts emerge in a world of relative facts and discuss their respective roles in connecting quantum theory and the world. The distinction between relative and stable facts resolves the difficulties pointed out by the no-go theorem of Frauchiger and Renner, and is consistent with the experimental violation of the Local Friendliness inequalities of Bong et al.. Basing the ontology of the theory on relative facts clarifies the role of decoherence in bringing about the classical world and solves the apparent incompatibility between the 'linear evolution' and 'projection' postulates.

## 1 Facts in Quantum Theory

The common textbook presentation of quantum theory assumes the existence of a classical world. A measurement involves an interaction between the classical world and a quantum system. The measurement produces a definite result, for instance a dot on a screen. The result is a fact by itself, but also establishes a fact about a quantum system. For instance, a certain measurement resulting in a definite record

✉ Andrea Di Biagio
andrea.dibiagio@uniroma1.it

Carlo Rovelli
rovelli@cpt.univ-mrs.fr

[1] Dipartimento di Fisica, Sapienza University of Rome, 00185 Rome, Italy

[2] Aix-Marseille Univ, Université de Toulon, CNRS, CPT, 13288 Marseille, France

[3] Perimeter Institute, 31 Caroline Street N, Waterloo, ON N2L 2Y5, Canada

[4] The Rotman Institute of Philosophy, 1151 Richmond St. N, London N6A5B7, Canada

establishes that at some time the $z$-component of the spin of an electron is $L_z = \frac{\hbar}{2}$. This is a fact.

Quantum probabilities are probabilities for facts, given other facts. The facts are therefore entries of which the probability amplitudes are function. In particular, facts are used as conditionals for computing probabilities of other facts. For instance, if the spin of the electron mentioned above is immediately measured in a direction at an angle $\theta$ from the $z$-axis, the probability to find the value $L_\theta = \frac{\hbar}{2}$ (a fact), given the fact that $L_z = \frac{\hbar}{2}$, is $P\left(L_\theta = \frac{\hbar}{2} | L_z = \frac{\hbar}{2}\right) = \cos^2\left(\frac{\theta}{2}\right)$. Hence, conditional probabilities of facts are what quantum mechanics is about.

Facts ascertained in a conventional measurement are *stable* in the following sense. If we know that one of $N$ mutually exclusive facts $a_i$ ($i = 1 \ldots N$) has happened, the probability $P(b)$ for another fact $b$ to happen is given by

$$P(b) = \sum_i P(b|a_i)P(a_i), \qquad (1)$$

where $P(a_i)$ is the probability that $a_i$ has happened and $P(b|a_i)$ is the probability for $b$ given $a_i$. We take Eq. (1) as a characterisation of *stable* facts.

This textbook presentation of quantum mechanics is incomplete because it assumes the existence of a classical world. An exactly classical world can exist only if current quantum theory has limited validity—for instance if physical collapse mechanisms exist [14, 23], or if quantum theory cannot be extended to systems with an infinite number of degrees of freedom [16], or other reasons. Quantum theory has however been universally successful so far, and there is no empirical evidence of its failure. This strongly suggest that real physical objects are classical (meaning they do not display quantum properties) only approximatively. There are no *exactly* classical objects, strictly speaking, as everything we interact with is made of atoms and photons, which obey quantum theory.

It is unconvincing to use concepts valid only within an approximation when formulating the fundamental theory of nature. Therefore the attempts at interpreting quantum theory as a universal theory do not rely on postulating classical objects. This is the case for instance for the Many Worlds interpretations [30, 33] and the de Broglie-Bohm pilot wave theory [11, 15]. Relational Quantum Mechanics (RQM) [21, 28, 29] is an interpretation of quantum theory as a universal theory, but it postulates *neither* classical objects *nor* unobservable worlds, unobservable variables or unobserved physics. Instead, RQM bases the interpretation of the theory on a larger ensemble of facts, of which stable facts are only a subset. These are called *relative* facts.

## 1.1 Relative Facts

Relative facts are *defined* to happen whenever a physical system interacts with another physical system. While relative facts play a central role in RQM, their definition and their usefulness are independent of the interpretation. We shall discuss this role in detail in the next section.

Let us consider two systems $\mathcal{S}$ and $\mathcal{F}$. If an interaction affects a variable $L_{\mathcal{F}}$ of $\mathcal{F}$ in a manner that depends on the value of a certain variable $L_{\mathcal{S}}$ of $\mathcal{S}$, then the value of $L_{\mathcal{S}}$ is a fact relative to $\mathcal{F}$. That is, whenever a system $\mathcal{F}$ is affected by a variable of another system, the value of that variable becomes a fact for $\mathcal{F}$. This is true by definition irrespectively of whether $\mathcal{F}$ is a classical system. The interaction with $\mathcal{F}$ is the context in which that variable takes a specific value; we call the system $\mathcal{F}$, in this role, a 'context'. The interaction with the context determines the fact that a certain variable has a value in that context.[1]

Stable facts are a strict subset of the relative facts: there are many relative facts that are not stable facts. Quantum theory provides probabilities relating relative facts, but these satisfy (1) only if $b$ and the $a_i$ are facts relative to the *same* system. That is, if we label facts with their context (writing $a^{(\mathcal{F})}$ for a fact relative to system $\mathcal{F}$), then it is always the case that

$$P(b^{(\mathcal{F})}) = \sum_i P(b^{(\mathcal{F})}|a_i^{(\mathcal{F})})P(a_i^{(\mathcal{F})}). \tag{2}$$

In contrast, whenever $\mathcal{W} \neq \mathcal{F}$, it is in general not the case that

$$P(b^{(\mathcal{W})}) = \sum_i P(b^{(\mathcal{W})}|a_i^{(\mathcal{F})})P(a_i^{(\mathcal{F})}). \tag{3}$$

If (3) holds, we say that the facts $a_i^{(\mathcal{F})}$ are *stable for* $\mathcal{W}$.

The failure of (3) is easily understood in terms of the standard language of quantum theory: it is the presence of interference effects. If $\mathcal{F}$ is sufficiently isolated it may be possible to maintain quantum coherence for the compound system $\mathcal{S} - \mathcal{F}$. The interaction entangles the two systems and interference effects between different values of the variable $L_{\mathcal{S}}$ can later be detected in the measurements by an observer $\mathcal{W}$. The probabilities for facts of the $\mathcal{S} - \mathcal{F}$ system relative to $\mathcal{W}$ can indeed be computed from an entangled state of the form

$$c_1|a_1\rangle \otimes |Fa_1\rangle + c_2|a_2\rangle \otimes |Fa_2\rangle, \tag{4}$$

where $a_i$ are values of $L_{\mathcal{S}}$ and $Fa_i$ are values of $\mathcal{F}$'s 'pointer variable' $L_{\mathcal{F}}$. Probabilities computed from this state violate (3) as they feature interference terms because what sums is amplitudes, not probabilities. The value of $L_{\mathcal{S}}$, therefore, is not a stable fact.

Hence facts relative to a system $\mathcal{F}$ cannot in general be taken as conditionals for computing probabilities of facts relative to a different system $\mathcal{W}$. Equation (1) holds only if $b$ and $a_i$ are facts relative to the same system, but fails in general if used for facts relative to different systems.

---

[1] We use 'variable' to denote any quantity that in the classical theory is a function on phase pace. We prefer to avoid the expression 'observable' because it is loaded with irrelevant extra baggage: the ideas of observation and a complex observer. The term 'context' is used here in a sense similar to that in [17]. However we do not require the context to be classical. See [1].

While the notation $\mathcal{S}$ for 'system', $\mathcal{F}$ for 'Friend' and $\mathcal{W}$ for 'Wigner' is meant to evoke the famous *Wigner's friend* thought experiment [34], in the discussion above there are no assumptions about the system $\mathcal{F}$ being quantum or classical, microscopic or macroscopic.

So, what exactly characterises a stable fact, among the relative facts? What gives rise to stable facts?

### 1.2 Decoherence

Since stability is a characteristic feature of the classical world, whose facts invariably satisfy (1), answering the questions above amounts to explaining in terms of relative facts what it takes for a system to be classical.

Various characterisations of a classical or semiclassical situation can be found in the literature: large quantum numbers, semiclassical wavepackets or coherent states, macroscopic systems, large or infinite number of degrees of freedom...All these features play a role in characterising classical systems in specific situations. But the key phenomenon that makes facts stable is decoherence [36, 39, 40]: the suppression of interference that happens when some information becomes inaccessible.

Consider two systems $\mathcal{F}$ and $\mathcal{E}$ ($\mathcal{E}$ for 'Environment'), and a variable $L_\mathcal{F}$ of the system $\mathcal{F}$. Let $Fa_i$ be the eigenvalues of $L_\mathcal{F}$. A generic state of the compound system $\mathcal{F} - \mathcal{E}$ can be written in the form

$$|\psi\rangle = \sum_i c_i |Fa_i\rangle \otimes |\psi_i\rangle, \tag{5}$$

where $|\psi_i\rangle$ are normalised states of $\mathcal{E}$. Let us define

$$\epsilon = \max_{i \neq j} |\langle \psi_i | \psi_j \rangle|^2. \tag{6}$$

Now, suppose that: (a) $\epsilon$ is vanishing or very small and (b) a system $\mathcal{W}$ does not interact with $\mathcal{E}$. Then the probability $P(b)$ of any possible fact relative to $\mathcal{W}$ resulting from an interaction between $\mathcal{F}$ and $\mathcal{W}$ can be computed from the density matrix obtained tracing over $\mathcal{E}$, that is,

$$\rho = \text{tr}_\mathcal{E} |\psi\rangle\langle\psi| = \sum_i |c_i|^2 |Fa_i\rangle\langle Fa_i| + O(\epsilon). \tag{7}$$

By posing $P(Fa_i^{(\mathcal{E})}) = |c_i|^2$, we can then write

$$P(b^{(\mathcal{W})}) = \sum_i P\left(b^{(\mathcal{W})} \Big| Fa_i^{(\mathcal{E})}\right) P\left(Fa_i^{(\mathcal{E})}\right) + O(\epsilon). \tag{8}$$

Thus, probabilities for facts $b$ relative to $\mathcal{W}$ calculated in terms of the possible values of $L_\mathcal{F}$ satisfy (3), up to a small deviation of order $\epsilon$. Hence the value of the variable $L_\mathcal{F}$ is a fact relative to $\mathcal{E}$ that is stable for $\mathcal{W}$ *to the extent* to which one ignores effects of order $\epsilon$. In the limit $\epsilon \to 0$, the variable $L_\mathcal{F}$ of the system $\mathcal{F}$ is exactly stable for $\mathcal{W}$.

Extensive theoretical work has shown that decoherence is practically unavoidable and extremely effective as soon as large numbers of degrees of freedom are involved [41]. The variables of $\mathcal{F}$ that decohere, namely the specific variables for which $\epsilon$ becomes small, are determined by the actual physical interactions between $\mathcal{F}$ and $\mathcal{E}$ (they are those variables that commute with the interaction Hamiltonian). The decoherence time, namely the time needed for $\epsilon$ to become so small that interference effects become undetectable by given observational methods, can be computed and is typically extremely short for macroscopic variables of macroscopic objects. All this is well understood.

It is important for what follows to emphasise two subtle aspects of decoherence.

First, decoherence is not an *absolute* phenomenon, but a relational one: it depends on how the third system $\mathcal{W}$ interacts with the combined system $\mathcal{F} - \mathcal{E}$. This is because assumption (b) above is just as crucial as assumption (a) in deriving (8). Another system $\mathcal{W}'$ that interacts differently with $\mathcal{F} - \mathcal{E}$ might be able to detect interference effects.

Second, decoherence implies that an event regarding two systems $\mathcal{F}$ and $\mathcal{E}$ is stable for at *third* system $\mathcal{W}$. Hence, a fact stable for $\mathcal{W}$ is not necessarily a fact relative to $\mathcal{W}$. That is, the variable $L_\mathcal{F}$ is stable for $\mathcal{W}$ even if the latter has not interacted with it, so there is no fact relative to $\mathcal{W}$ yet. This is what allows one to say that, with respect to $\mathcal{W}$, the "state of the system $\mathcal{F}$ has collapsed into the state $|Fa_i\rangle$ with probability $P(Fa_i) = |c_i|^2$," even though $\mathcal{W}$ has not interacted with $\mathcal{F}$.

These observations show that decoherence does not imply that there is a perfectly classical world of absolute facts, although it does explain why (and when) we can reason in terms of stable, hence approximatively classical, facts.[2]

### 1.3 Measurements

If two systems $\mathcal{S}$ and $\mathcal{F}$ interact so that their respective variables $L_\mathcal{S}$ and $L_\mathcal{F}$ get entangled, and if $L_\mathcal{F}$ is stable for $\mathcal{W}$, it follows immediately from the definitions that also $L_\mathcal{S}$ is stable for $\mathcal{W}$.

This is precisely what happens in a typical quantum measurement of a variable $L_\mathcal{S}$ in a laboratory. Thinking of $\mathcal{S}$, $\mathcal{F}$ and $\mathcal{W}$ as, respectively, the system being measured, the apparatus and the experimenter, we can separate the measurement in three stages:

1. An interaction between the system and the apparatus entangles $L_\mathcal{S}$ with a pointer variable $L_\mathcal{F}$ of the apparatus.

---

[2] There is a limit case in which a fact can be stable even in the absence of decoherence. This is when one of the amplitudes in (5), say $c_1$, has absolute value close to 1. If $\mathcal{W}$ does not interact with $\mathcal{E}$, then the probabilities for facts relative to $\mathcal{W}$ can be computed using $\rho' = |Fa_1\rangle\langle Fa_1| + O(\eta)$, where $\eta = 1 - |c_1|$. In this case, one can reason as if the value of $L_\mathcal{F}$ were a fact relative to $\mathcal{W}$, up to order $\eta$ effects. Thus, when a fact relative to a system has very high probability, then it is stable for other systems, because the interference effects are small. Notice the differences with the Einstein-Podolsky-Rosen criterion for an 'element of reality:' the above only holds to the extent to which $\mathcal{W}$ cannot interact with $\mathcal{E}$, and there needs to be a fact relative to $\mathcal{E}$ in the first place.

2. $L_\mathcal{F}$ gets correlated with a large number of microscopic variables (forming $\mathcal{E}$) that are inaccessible to the observer $\mathcal{W}$.
3. The observer $\mathcal{W}$ interacts with the pointer variable $L_\mathcal{F}$ to learn about $L_\mathcal{S}$.

Let's trace this same story in terms of relative facts:

1. A relative fact is established between $\mathcal{S}$ and $\mathcal{F}$.
2. A relative fact is established between $\mathcal{F}$ and $\mathcal{E}$. Since $\mathcal{W}$ does not interact with $\mathcal{E}$, this stabilises the previous fact for $\mathcal{W}$.
3. A relative fact is established between $\mathcal{F}$ and $\mathcal{W}$. This has consequences on $\mathcal{W}$'s future interactions with $\mathcal{S} - \mathcal{F}$.

Already at stage 2, the observer can apply (3) since the interaction with the inaccessible degrees of freedom greatly suppresses interference terms. The observer might say "$L_\mathcal{S}$ has been measured," and assume that the pointer of the apparatus moved one way or the other. In the mathematical formalism, $\mathcal{W}$ can assume that "$\mathcal{S}$'s wavefunction has collapsed." Note however that neither the value of $L_\mathcal{S}$ nor that of $L_\mathcal{F}$ is a fact for $\mathcal{W}$ at this stage. Stability simply allows $\mathcal{W}$ to "de-label" facts relative to $\mathcal{F}$.

It is is only at stage 3 that the value of $L_\mathcal{F}$ becomes a fact for $\mathcal{W}$. Note that the value of $L_\mathcal{S}$ is still not a fact relative to $\mathcal{W}$, it is merely a stable fact for $\mathcal{W}$. However, based on the value of $L_\mathcal{F}$, the experimenter $\mathcal{W}$ can update the state for $\mathcal{S}$. The experimenter can reason *as if* $L_\mathcal{S}$ took the value that she read on the apparatus' pointer variable.

It is the way that $\mathcal{W}$, $\mathcal{F}$ and $\mathcal{E}$ couple to each other that makes $\mathcal{F}$ a measuring apparatus for $\mathcal{W}$. The stability of $\mathcal{F}$ for $\mathcal{W}$ extends to all other variables that interact with $\mathcal{F}$, hence $\mathcal{W}$, applying quantum mechanics, might say that "$\mathcal{F}$ causes $\mathcal{S}$ to collapse." But, in fact, another system $\mathcal{W}'$ that couples differently to these systems might still be able to detect interference effects. Indeed this is what happens in the Wigner's friend scenarios.

In summary, we can distinguish two notions of facts that play a role in quantum theory: relative facts and stable facts. Quantum theory allows us to talk about relative facts and compute probabilities for them. Equation (2) holds but (3) does not. The violation of (3) is quantum interference.

Stable facts are a subset of the relative facts. They satisfy (3). A relative fact about a system $\mathcal{F}$ is stable for a system $\mathcal{W}$ if $\mathcal{W}$ has no access to a system $\mathcal{E}$ that is sufficiently entangled with $\mathcal{F}$. Stability is only approximate and relational. Approximate, because no fact is exactly stable for any finite $\epsilon$. Relational, because it depends on how the putative 'observer' system couples to the system and the environment.

## 2 Facts and Reality

We have given definitions of relative and stable facts, and studied their properties. In this section we discuss the roles of relative and stable facts for the interpretation of quantum theory, namely for the relation between the formalism and the reality it describes.

### 2.1 The Link Between the Theory the World

Let us compare advantages and difficulties of interpreting either stable or relative facts as the link between theory and reality.

Stable facts are taken as the link between the formalism and the world in textbook interpretations of quantum theory. They are the conventional 'measurement outcomes' in a macroscopic laboratory. They are similar to the facts of classical mechanics because, in the world described by classical mechanics, all facts (variables having certain values at certain times) are exactly stable: the (epistemic) probabilities for them to happen are *always* exactly consistent with (1). In quantum mechanics, facts stable with respect to us are ubiquitous because of the ubiquity of decoherence.

There are however two difficulties in taking stable facts as the basis of the quantum ontology. First, stability is relational. Facts are stable only for a system that does not have sufficiently precise interactions with an environment system. The system and environment are still in a superposition with respect to a third system. Therefore one does not avoid relationalism by restricting to stable facts. Second, more seriously, stability is only approximate in general. At no point the interference terms perfectly vanish.

These are serious difficulties if we want to take stable facts as the only primary elements of reality. How stable does a fact need to be before it is real? And with respect to what systems does it have to be stable, in order to be real? Any answer to these questions is bound to be as unsatisfactory as the textbook interpretation that requires a classical world. The alternative is to embrace the contextuality of the theory in full, and base its ontology on all relative facts.

Relative facts form the basis of a realist interpretation in Relational Quantum Mechanics (RQM) [21, 28, 29]. The fundamental contextuality that characterises quantum theory is interpreted in RQM as the discovery that facts about a system are always defined relative to another system, with which the first system interacts.

In the early history of quantum theory it was recognised that every measurement involves an interaction, and it was said that variables take values only upon measurement. RQM notices that every interaction is in a sense a measurement, in that it results in the value of a variable to become a fact. These facts are not absolute, they belong to a context. And there is no 'special context': any system can be a context for any other system.

The quantum state ('the wavefunction') does not have an ontic interpretation in RQM. The state is not a 'thing', nor a condition of a system. Rather, it is what a physicist uses to calculate probabilities for relative facts between physical systems to happen, given the relevant information she has. It follows that RQM has no use for a 'wavefunction of the universe' that forever evolves unitarily, as this would be a tool to calculate probabilities of facts relative to something that does not exist: a system that is not part of the universe.

Unlike other epistemic interpretations of quantum theory [7, 13, 18, 32, 37], the ontology of RQM is realist (as is the formal reconstruction theorem it inspired [19]). Realist in the sense that it is not about agents, beliefs, observers, or experiences: it is about real facts of the world and relative probabilities of their occurrence. The ontology is relational, in the sense that it is based on facts established at interactions and are labelled by physical contexts. Relative facts, therefore, provide a relational but realist interpretation to quantum theory which does not need to refer to complex agents.

## 2.2 No-Go Theorems for Absolute Facts

A number of results have recently appeared in the literature as no-go theorems for absolute (non-relative) facts [5, 6, 12]. These results analyse the extended Wigner's friend scenario (EWFS), where instead of a superobserver reasoning about his friend in a hermetically sealed lab, there are two superobservers each reasoning about their own friend, with the friends entangled.

### 2.2.1 Quantum Theory Cannot Consistently Describe the use of Itself

In [12], Frauchiger and Renner use a EWFS to show that quantum theory is inconsistent under a certain number of assumptions. A key assumption used to derive the contradiction is the absolute nature of facts. This is assumption (C) in the paper, which can be stated as follows: "If $\mathcal{W}$, applying quantum theory, concludes that $\mathcal{F}$ knows that $L_S = a$, then $\mathcal{W}$ can conclude that $L_S = a$." The 'C' stands for consistency: the authors argue that this assumption is required to deem the theory consistent: different agents using the same theory must arrive at the same conclusions. This is an ironic choice of name from the point of view of the present discussion, as it is precisely this assumption that leads to contradiction, according to RQM.

In terms of relative facts, assumption (C) implies: "If $\mathcal{W}$, applying quantum theory, can be certain that $L_S = a$ relative to $\mathcal{F}$, then $\mathcal{W}$ can reason as if $L_S = a$ also relative to $\mathcal{W}$." Now, as we have shown, this holds only if every fact relative to $\mathcal{F}$ is stable for $\mathcal{W}$, which is not a given and depends on the physics. Therefore Assumption (C) only holds if $\mathcal{S}$ or $\mathcal{F}$ decohere with respect to $\mathcal{W}$. In the Frauchiger and Renner protocol, the superobservers are supposed to have full quantum control on their friends and the contents of their labs. Thus, by definition, what is stable for the friends is not stable for $\mathcal{W}$. Hence the contradiction follows from inappropriately mixing contexts: forgetting that facts are relative and therefore (3) does not hold in general.

Indeed, as pointed out in [38] and worked out in detail in [26, 27], no contradiction can be derived if one additionally assumes that what is decoherent for the friends (the lab they are in) is also decoherent for $\mathcal{W}$. The analysis of how an agent should reason about an experiment that will be performed on him has not been done within RQM yet. The reader is invited to consider the analysis within QBism [9, 10], since also QBism holds that assumption (C) fails and quantum states are only used to calculate probabilities from the point of view of a given system.

### 2.2.2 A Strong No-Go Theorem on the Wigner's Friend Paradox

Another enlightening result is the recent Bong et al. [5], which is a strengthening of a previous result by Brukner [6, 25]. In [5], the authors show that the conjunction of (a) absoluteness of observed events, (b) no superdeterminism and (c) locality imply that correlations in the extended Wigner's friend scenario must satisfy some inequalities, called the Local Friendliness (LF) inequalities. Like Bell's inequalities [2, 3], these are derived in a theory-independent way. The authors then show that quantum theory predicts the violation of these inequalities. Thus the universal validity of quantum theory implies that one of the three properties above does not hold.

The word 'locality' means different things in different physics communities. The notion used to derive the LF inequalities is the one that Bell used to derive his inequalities in [2]. In operational language, (c) says that a free choice does not alter the probabilities of a spacelike separated event. Most epistemic interpretations accept this notion of locality [8], while it is rejected by pilot-wave interpretations [15]. No superdeterminism simply means that free choices are possible so that, in particular, the measurement settings can be chosen so as to be uncorrelated with other relevant variables. See [35] for an in depth analysis on the notions of locality and superdeterminism in the context of the implications of Bell's theorems.

If one believes that quantum theory holds at arbitrary scales, wishes to maintain locality and reject superdeterminism, one has no choice but to reject the absoluteness of observed events. Absoluteness of observed events means that if $\mathcal{W}$ deems that $\mathcal{F}$ is an observer, then $\mathcal{W}$ can use (3)—even if $\mathcal{W}$ has full quantum control on $\mathcal{F}$. This clearly does not hold in RQM: if $\mathcal{W}$ has full quantum control on $\mathcal{F}$ then facts relative to $\mathcal{F}$ are not stable for $\mathcal{W}$ and thus (3) does not hold. Note that in RQM there are no special 'observer' systems so an 'observed event' is simply a fact relative to a given system.

Remarkably, the LF inequalities have already been experimentally violated when the friends are single photons [5]. One might be tempted to dismiss the results on the ground that photons do not generate facts ("photons are not observers"), but this opens the problem of deciding which systems give rise to facts. If quantum theory is universally valid, advances in quantum technologies will allow to perform the same experiment with increasingly complex 'friends'. The predictions of quantum theory remain the same: the statistics are incompatible with the assumptions of absolute facts.

The violation of the LF inequalities is no way in tension with the relational interpretation. The opposite is true: the result is taken as evidence that the facts quantum theory deals with are facts relative to systems.[3]

---

[3] For a discussion of the implications of the ontology of RQM to Bell's inequalities, see [20, 31] and [22, 24].

## 2.3 Conclusions and Final Comments

The insight of Relational Quantum Mechanics (RQM) is that recognising the relative nature of facts offers a straightforward solution to the measurement problem. The measurement problem is the apparent incompatibility between two postulates: the 'projection' and the 'linear evolution' postulate. Both postulates can be correct: they refer to facts relative to different systems. Say that $\mathcal{S}$ interacts with $\mathcal{F}$, so that a fact relative to $\mathcal{F}$ is established. Then the projection postulate is used to update the state of $\mathcal{S}$ with respect to $\mathcal{F}$, while the unitary evolution postulate is used to update the state of $\mathcal{S} - \mathcal{F}$ with respect to a third system $\mathcal{W}$.

In a slogan: 'Wigner's facts are not necessarily his Friend's facts'.

This by no means implies that when Wigner and his friend compare notes they find contradictions [28]. Interactions between $\mathcal{S}$ and $\mathcal{F}$ do have influence on the facts relative to $\mathcal{W}$. Indeed, after an interaction, $\mathcal{S}$ and $\mathcal{F}$ are entangled relative to $\mathcal{W}$, meaning that in interacting with the two systems, $\mathcal{W}$ will find the two correlated. Therefore Wigner will always agree with his Friend about the value of $L_\mathcal{S}$ once he too interacts with them. In this sense, relative facts correspond to real events, they have universal empirical consequences.

Still, accepting the relativity of all facts is a strong conceptual step. It amounts giving up the absolute nature of facts, namely, the existence of an absolute 'macroreality' in the language used in discussions of Bell's inequalities [35]. Such a macroreality only emerges approximately, relative to systems for which decoherence is sufficiently strong.

Decoherence has always played a peculiar role in the discussions on the measurement problem. On the one hand, it is simply a true physical phenomenon, obviously relevant for shedding light on quantum measurement. On the other hand, there is consensus that decoherence alone is not a solution of the measurement problem, because it does not suffice to provide a link between theory and reality. Decoherence needs an ontology. Relative facts provide such a general ontology that is well defined with or without decoherence. Decoherence clarifies why a large class of relative facts are stable for us and thus form the stable classical world we live in.

The violation of (1) when used for facts relative to different systems sheds also some light on the underpinnings of quantum logic. The violation of (1), indeed, has been interpreted as a violation of classical logic [4], as it can be written as

$$P\big(b \text{ and } (a_1 \text{ or } a_2)\big) \neq P\big((b \text{ and } a_1) \text{ or } (b \text{ and } a_2)\big), \tag{9}$$

in contradiction with the classical logic theorem

$$b \text{ and } (a_1 \text{ or } a_2) = (b \text{ and } a_1) \text{ or } (b \text{ and } a_2). \tag{10}$$

The apparent violation of logic is understood in RQM as a result of forgetting that facts are relative: labelled by a context, as Bohr has repeatedly pointed out. Facts relative to a context cannot be used, in general, to compute probabilities of facts related to other contexts because what is a fact in a certain context is not necessarily a fact in other contexts.

As a final remark, observe that if the quantum state has no ontic interpretation, the only meaning of 'being in a quantum superposition' is that interference effects are to be expected. Saying "Friend is in a quantum superposition" does not mean anything more than saying that Wigner would be mistaken in using (3). It has no implications on how Friend would "feel" while being in a superposition. Friend sees a definite result of her measurement, a fact, and this does not prevent Wigner from having the chance to see an interference effect in *his* facts. Wigner's friend does not stop being an observer simply because Wigner has a chance to detect interference effects in *his* facts. Schrödinger's cat has no reason to feel 'superposed'.

**Acknowledgements** CR thanks Cristian Mariani, Alexia Auffèves, Philippe Grangier and all the participants at the reading group of the Institut Néel in Grenoble, for very stimulating discussions at the origin of this paper. Both authors wish to thank Eric Wiseman for a clarification on the significance of the work [5] and Axel Polaczek for useful comments on a first version of this note. This work was made possible through the support of the FQXi Grant FQXi-RFP-1818 and of the ID# 61466 grant from the John Templeton Foundation, as part of the "The Quantum Information Structure of Spacetime (QISS)" Project (www.qiss.fr).

**Funding** Open access funding provided by Università degli Studi di Roma La Sapienza within the CRUI-CARE Agreement.